\documentclass[%
 reprint,
superscriptaddress,
 amsmath,amssymb,
 aps,
prl
]{revtex4-1}
\usepackage[dvipsnames]{xcolor}
\usepackage[english]{babel}
\usepackage{hyperref}
 
\makeatletter
\def\bbl@set@language#1{%
  \edef\languagename{%
    \ifnum\escapechar=\expandafter`\string#1\@empty
    \else\string#1\@empty\fi}%
  \@ifundefined{babel@language@alias@\languagename}{}{%
    \edef\languagename{\@nameuse{babel@language@alias@\languagename}}%
  }%
  \select@language{\languagename}%
  \expandafter\ifx\csname date\languagename\endcsname\relax\else
    \if@filesw
      \protected@write\@auxout{}{\string\select@language{\languagename}}%
      \bbl@for\bbl@tempa\BabelContentsFiles{%
        \addtocontents{\bbl@tempa}{\xstring\select@language{\languagename}}}%
      \bbl@usehooks{write}{}%
    \fi
  \fi}
\newcommand{\DeclareLanguageAlias}[2]{%
  \global\@namedef{babel@language@alias@#1}{#2}%
}
\makeatother

\DeclareLanguageAlias{en}{english}

\usepackage{graphicx}% Include figure files
\usepackage{amsmath}
\usepackage{array}
\usepackage{amssymb}
\usepackage{dcolumn}% Align table columns on decimal point
\usepackage{bm}% bold math
\let\savecorresponds\corresponds
\let\corresponds\relax
\usepackage{mathabx}
\usepackage{enumitem} 
\usepackage{tabularx}
\usepackage{soul}
\usepackage{ulem}
\let\corresponds\savecorresponds
\renewcommand{\vec}[1]{\mathbf{#1}}

\def\vec#1{\boldsymbol{#1}}

\def\pd2v#1#2#3{\frac{\partial^2 #1}{\partial #2 \partial #3}}

\def \vec#1{\mathbf{#1}}
\def \2x2mat#1#2#3#4{
\left( \begin{array}{cc}
#1 &  #2 \\  #3 &  #4
\end{array} \right)
}

\begin{document}

\preprint{APS/123-QED}

\title{Arbitrary spatial mode sorting in a multimode fiber}%  \\

\author{Hugo Defienne}
\affiliation{School of Physics and Astronomy, University of Glasgow, Glasgow G12 8QQ, UK\
}%
\author{Daniele Faccio}%
\affiliation{School of Physics and Astronomy, University of Glasgow, Glasgow G12 8QQ, UK\
}%

\date{\today}
\begin{abstract}
Sorting spatial optical modes is a key challenge that underpins many applications from super-resolved imaging to high-dimensional quantum key distribution. However, to date implementations of optical mode sorters only operate on specific sets of modes, such as those carrying orbital angular momentum, and therefore lack versatility with respect to operation with an arbitrary spatial basis. Here, we demonstrate an arbitrary spatial mode sorter by harnessing the random mode mixing process occurring during light propagation in a multimode fibre by wavefront shaping. By measuring the transmission matrix of the fibre, we show sorting of up to $25$ transverse spatial modes of the Fourier, Laguerre-Gaussian and a random basis to an arbitrary set of positions at the output. Our approach provides a spatial mode sorter that is compact, easy-to-fabricate, programmable and usable with any spatial basis, which is promising for quantum and classical information science. 
\end{abstract}

\maketitle

A spatial mode sorter transforms a given spatial mode to a specific position in a transverse plane. Such a device is typically used to decompose a complex input optical signal into a specific spatial basis. One of the simplest examples is a convergent lens, that uniquely distributes the Fourier components of incoming light across different positions in the lens focal plane. In recent years, the development of mode sorting devices has attracted much attention because of the potential that transverse spatial modes (and knowledge of how these compose a given signal) hold for implementing fundamental optical tasks~\cite{rubinsztein-dunlop_roadmap_2017}. In classical optics, decomposing an image in the Hermite-Gaussian (HG) basis enables for example to improve image spatial resolution~\cite{tsang_quantum_2016,zhou_quantum-limited_2019}, and the use of a Laguerre-Gaussian (LG) basis for spatial multiplexing allows to increase the capacity of optical communication systems~\cite{wang_terabit_2012,yan_high-capacity_2014,willner_optical_2015}. In quantum optics, transverse spatial modes are used for producing high-dimensional quantum states~\cite{mair_entanglement_2001,fickler_quantum_2012} that hold potential for quantum computing and simulation~\cite{cardano_quantum_2015,brandt_high-dimensional_2020}, communication~\cite{mirhosseini_high-dimensional_2015,cozzolino_high-dimensional_2019} and fundamental studies~\cite{malik_multi-photon_2016}. 

However, even if some technologies for manipulating spatial modes of light are  commercially available and widely used~\cite{forbes_creation_2016}, spatial mode sorting techniques are still at their early development stage. Among them, \textit{phase-flattening} is a well established scheme that was originally introduced to sort LG-modes of different orbital angular momentum (OAM)~\cite{mair_entanglement_2001}. This approach has the advantage of being simple to implement because it only consists of a spatial light modulator (SLM) and a single mode fibre, but it also has drawbacks~\cite{qassim_limitations_2014} including that it requires to perform $d$-projective measurements over time ($d$ is the number of sorted modes) and is restricted to specific families of modes. This technique was recently extended to LG-modes with different radial index~\cite{bouchard_measuring_2018-3} and HG-modes~\cite{hiekkamaki_near-perfect_2019}, but still measuring projections over time. More recently, \textit{full-field mode sorting} systems (i.e. no projective measurements) were developed for decomposing light into LG-modes. Examples range from systems using fixed diffractive optical elements~\cite{morizur_programmable_2010,lavery_refractive_2012,ruffato_compact_2018} to those based on multiple phase screens programmed with SLMs~\cite{berkhout_efficient_2010,mirhosseini_efficient_2013,fontaine_laguerre-gaussian_2019,fickler_full-field_2020}. Nevertheless, these systems are currently restricted to LG-modes and are challenging to implement because they require light to be reflected by a large number of phase screens for efficient sorting. This number scales as $6d+1$ in the case of sorting $d$ modes between arbitrary spatial modes~\cite{lopez-pastor_arbitrary_2019}. Finally, we also note that scattering in a layer of paint has been exploited for mode sorting~\cite{fickler_custom-tailored_2017} by using time-consuming optimisation-based wavefront shaping approaches~\cite{vellekoop_focusing_2007}.

\begin{figure}
\includegraphics[width=1 \columnwidth]{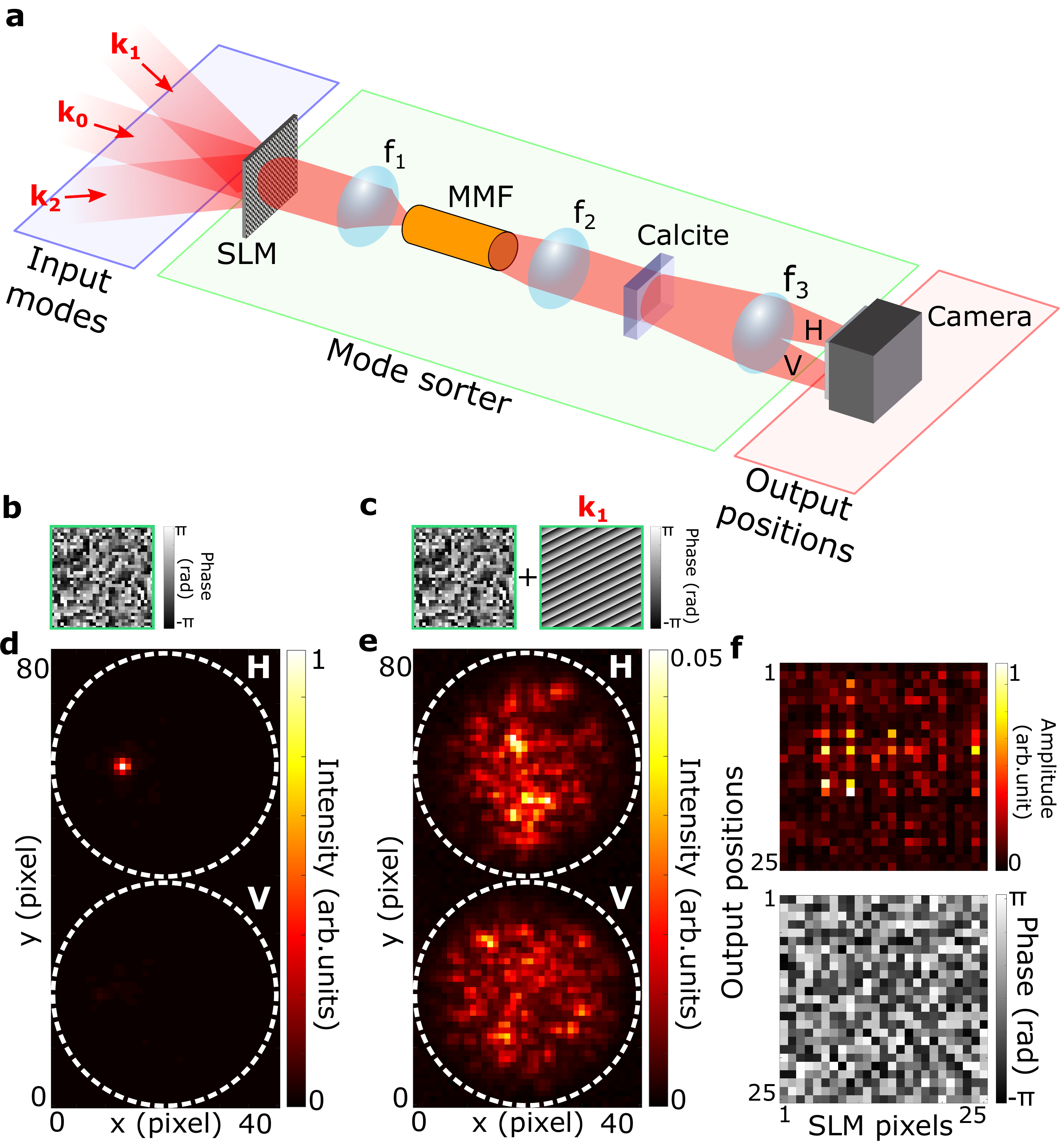} % this command will be ignored
\caption{\label{Figure1}  \textbf{a}, A phase-only spatial light modulator (SLM) shapes and injects monochromatic polarised light ($810$nm) into a $5$cm long $50\mu$m core diameter graded index multimode fiber (MMF) using a lens $f_1=20$mm. The output surface of the fibre is imaged on the camera by two lenses $f_2=20$mm and $f3=200$mm. A calcite is used to produce two vertical (V) and horizontal (H) polarised images next to each other. The mode sorter consists of the SLM and the MMF. Input modes are the transverse spatial modes of light illuminating the SLM and output positions are the camera pixels. \textbf{b}, Phase mask programmed on the SLM to focus light at normal incidence $\vec{k_0}$. \textbf{c}, A phase ramp corresponding to an input mode $\vec{k_1} \neq \vec{k_0}$ is superimposed onto the focusing SLM mask. \textbf{d} and \textbf{e}, Intensity images acquired under $\vec{k_0}$ and $\vec{k_1}$ illuminations, respectively. \textbf{f}, Amplitude and phase of a $25 \times 25$ subset of the measured transmission matrix (TM).   }
\end{figure}

Here, we implement a simple full-field mode sorting system that can operate on any basis. For this, we leverage the complex spatial mode mixing process performed by a multimode fibre (MMF)  by using a transmission matrix (TM) based wavefront shaping technique. The optical TM was introduced by Popoff et al~\cite{popoff_measuring_2010} for manipulating monochromatic light through a layer of paint and was then extended to other complex systems such as MMFs~\cite{carpenter_110x110_2014,ploschner_seeing_2015-3} and can also work with light sources including optical pulses~\cite{mounaix_spatiotemporal_2016,mounaix_control_2019} and photon-pairs~\cite{defienne_two-photon_2016}. Recently, the TM was also used to design complex linear optical networks for classical~\cite{matthes_optical_2019} and quantum~\cite{Leedumrongwatthanakun_programmable_2020} simulations. In our work, we extend the range of applications to spatial mode sorting. Using the TM of a MMF, we report experimental and simulated results of sorting up to $25$ modes and analyse the performance of our approach with examples taken from the  Fourier basis, the LG basis and a random basis. 

\begin{figure*}
\includegraphics[width=1 \textwidth]{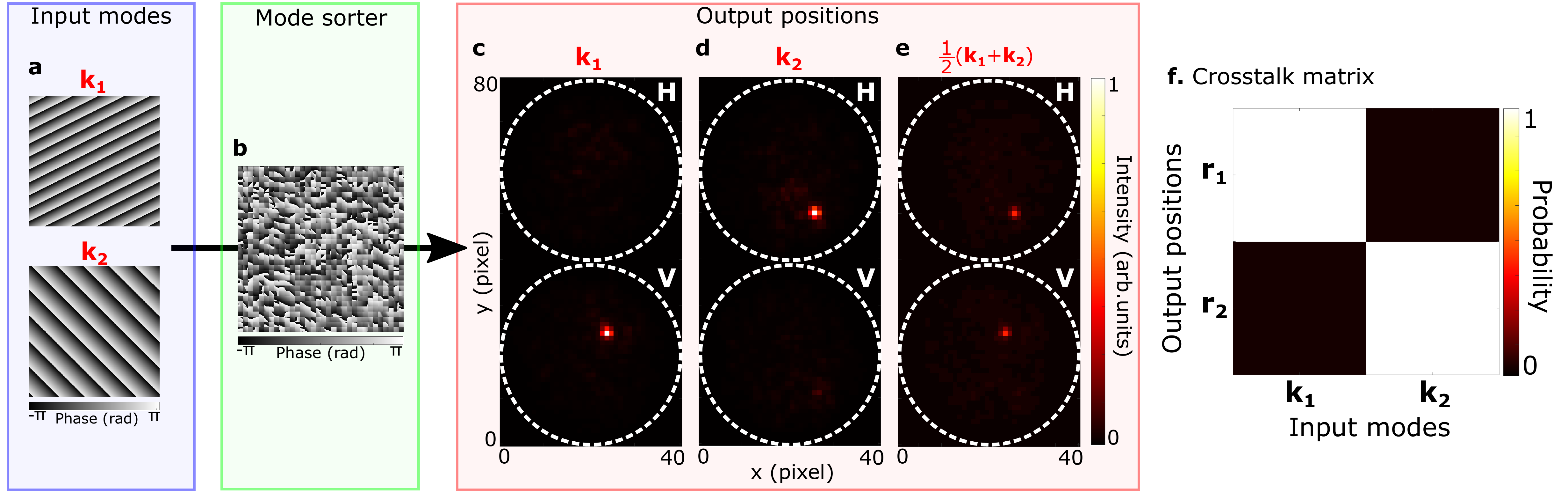} % this command will be ignored
\caption{\label{Figure2} \textbf{a}, Spatial phase components of two input modes $\vec{k_1}$ and $\vec{k_2}$ of the Fourier basis. \textbf{b}, Phase mask programmed on the SLM to implement a two-dimensional mode sorter $\vec{k_1} \rightarrow \vec{r_1}$ and $\vec{k_2} \rightarrow \vec{r_2}$ in the MMF. \textbf{c,d} and \textbf{e}, Intensity images measured for input mode $\vec{k_1}$, input mode $\vec{k_2}$ and a linear combination of them $1/2(\vec{k_1}+\vec{k_2})$, respectively. Light is focused in two different camera positions denoted $\vec{r_1}$ and $\vec{r_2}$. \textbf{f}, Crosstalk matrix of the programmed mode sorter showing a sorting ability of $97.5(1)\%$.}
\end{figure*}

Figure~\ref{Figure1}.a describes an experimental setup composed of an SLM that injects structured light into a MMF and a camera that measures the output speckle images in both polarisations. The TM of the MMF ($T$) is measured by illuminating the SLM at normal incidence with a collimated Gaussian beam (input mode $\vec{k_0}$) and using a co-propagating reference, as detailed in~\cite{popoff_measuring_2010}. $T$ is a complex matrix that links optical fields between $N=32\times32$ SLM macropixels and $M=80 \times 40$ camera pixels (Fig.~\ref{Figure1}.f). One of the most basic tasks that the TM can achieve is to focus light through the MMF. Using the complex conjugate operator $T^\dagger$, an SLM phase mask is calculated and programmed (Fig.~\ref{Figure1}.b) to focus scattered light at a targeted camera pixel~\cite{popoff_measuring_2010}, as shown in the output intensity image in Figure~\ref{Figure2}.d. Interestingly, focusing light using the TM can be seen as a very simple one-dimensional mode sorting operation: light from a input mode $\vec{k_0}$ is directed to a specific position in the camera plane. If a mode with a different wave-vector $\vec{k_1} \neq \vec{k_0}$ is inserted at the input, which is done experimentally by superimposing a phase ramp on the focusing phase mask on the SLM (Fig.~\ref{Figure2}.c), the focusing effect at the output is lost and the mode is not sorted (Fig.~\ref{Figure1}.e). 

We build our TM based mode sorting approach based on this method to focus light through the MMF. First, we arbitrarily choose spatial modes within a given spatial mode basis. In the example detailed in Figure~\ref{Figure2}, we selected two modes from the Fourier basis characterised by wave-vectors $\vec{k_1}$ and $\vec{k_2}$ ($\neq \vec{k_0}$). The basis is represented by a change of basis matrix $P$ in which each column is a complex vector listing all components of the corresponding mode written in the SLM plane position basis (see Supplementary Information). Second, we select two positions $\vec{r_1}$ and $\vec{r_2}$ within the illuminated area on the camera and define a target mode sorting operator $M$. $M$ is a real matrix linking input modes (column) to output positions (lines). In order to implement the sorting operation $\vec{k_1} \rightarrow \vec{r_1}$ and $\vec{k_2} \rightarrow \vec{r_2}$, $M$ is written as a matrix composed of zeros with only two ones located at the crossing between the column associated with $\vec{k_1}$ and the line associated with $\vec{r_1}$, and the column $\vec{k_2}$ and line $\vec{r_2}$. Finally, the phase mask that we program on the SLM for implementing the mode sorting operation (Fig.~\ref{Figure2}.b) is calculated using the formula~\cite{matthes_optical_2019}
\begin{equation}
\label{equ1}
\vec{\Phi} = \arg \left[ \mbox{diag} \left( T^\dagger M P^\dagger \right) \right]
\end{equation}
where $\vec{\Phi}$ is a vector associated with the phase mask, \textit{diag} refers to the diagonal of the matrix and $\textit{arg}$ to is the complex argument.

The physics underlying Eq.~\eqref{equ1} can be understood when considering the propagation of the input field through the MMF. Let's first consider an ideal situation in which the SLM is replaced by an optical system that can perform the linear operation $T^\dagger M P^\dagger$, where $M$ and $P$ represent arbitrary target and change-of-basis matrices. The output field $E^{out}$ obtained after propagation of an incoming field $E^{in}$ through the MMF is then written
\begin{equation}
\label{equ2}
E^{out} = T \left [ T^\dagger M P^\dagger \right] E^{in} \approx  M P^\dagger E^{in} 
\end{equation}
For mode sorting, $M$ can be written as an identity matrix and Eq.~\eqref{equ2} then describes a change of basis operation between an arbitrary spatial basis and output spatial positions, namely an arbitrary mode sorting process. Note that the approximation used in Eq.~\eqref{equ2} directly relies on the complex spatial mode mixing process performed by the MMF. Indeed, as shown experimentally in Fig.~\ref{Figure1}.f, a subset of the TM measured in the SLM and camera pixel basis can be approximated by a random complex matrix~\cite{popoff_measuring_2010,ploschner_seeing_2015-3}. In this case, one may write $T T^\dagger =  1\!\!1 + H / \sqrt{N} $ where $N$ is the number of columns of $T$ and $H$ is a random matrix of complex coefficients with unity variance (see SI). Equation~\eqref{equ2} is then only valid for $N \gg 1$, which is the case in our experiment ($N=1024$). However, in a realistic situation, an SLM can only shape the phase of the field in a specific optical plane, which means that it only controls the phase components of the diagonal coefficients of $T^\dagger M P^\dagger$. Such practical limitations effectively reduce the number of degrees of control from $2N^2$ (phase and amplitude $N\times N$ matrix coefficients) to $N$ (phase components of an optical plane), which has the consequence of decreasing the overall efficiency of the mode sorter compared to the ideal case. 

To test our mode sorter, we measured intensities at the output for input modes $\vec{k_1}$ and $\vec{k_2}$. In the experiment, input modes are generated by superimposing their corresponding phase masks on top of the mode sorting phase mask on the SLM. Figures~\ref{Figure2}.c and d show that light focuses at the two targeted positions $\vec{r_1}$ and $\vec{r_2}$ when either of the modes $\vec{k_1}$ and $\vec{k_2}$ are inserted at the input, respectively. Moreover, Figure~\ref{Figure2}.e shows that light focuses to both positions simultaneously when a linear combination of modes $1/2(\vec{k_1}+\vec{k_2})$ is programmed at the input. The mode sorting operation is characterised by a cross-talk matrix, shown in Figure~\ref{Figure2}.c. An average sorting ability $\bar{p}$ of $97.5(1)\%$ is calculated from the crosstalk matrix coefficients $I_{nk}$ ($n^{th}$ line and $k^{th}$ column) using the formula $\bar{p}=\sum_{n=1}^d \frac{I_{nn}}{\sum_{k=1}^d{I_{kn}}}$~\cite{fickler_custom-tailored_2017} (see SI).

\begin{figure}
\includegraphics[width=1 \columnwidth]{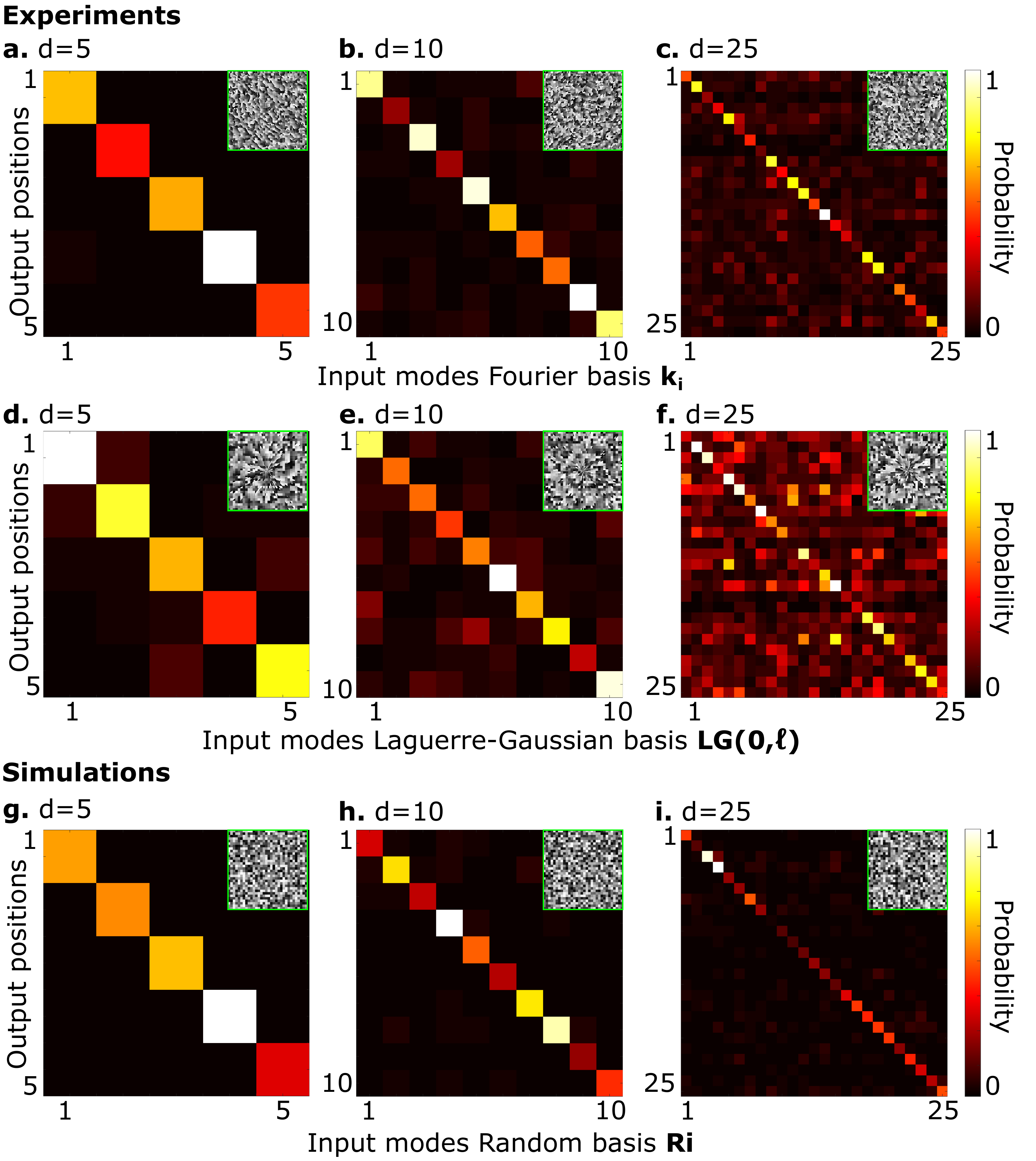} % this command will be ignored
\caption{\label{Figure3} \textbf{a},\textbf{b} and \textbf{c}, Experimental results of mode sorting in the Fourier basis $\vec{k_i}$ ($i \in [\![1;25]\!]$) with $d=5$, $d=10$ and $d=25$ modes. Average sorting ability are $93(3)\%$, $70(9)\%$ and $25(10)\%$, respectively. Inset images show the corresponding phase masks programmed on the SLM. \textbf{d},\textbf{e} and \textbf{f}, Experimental results of sorting Laguerre-Gaussian (LG) modes with radial number $p=0$ and azimutal number $\ell \in [\![-12;12]\!]$ using $d=5$, $d=10$ and $d=25$ modes. Average sorting ability are $82(3)\%$, $56(7)\%$ and $15(6)\%$, respectively. Insets show the SLM phase masks programmed in each case. \textbf{g},\textbf{h} and \textbf{i}, Results of mode sorting in a random basis $\vec{R_i}$ ($i \in [\![1;25]\!]$) obtained by simulating light propagating through the MMF with an experimentally measured TM for $d=5$, $d=10$ and $d=25$ modes. Average sorting ability are $97(3)\%$, $83(6) \%$ and $45(17) \%$, respectively.}
\end{figure}

Figure~\ref{Figure3} shows results of spatial mode sorting involving up to $25$  modes from different spatial basis sets. Using the same TM from the previous experiment, we calculated new phase masks with Eq.~\ref{equ1}, for three cases of  sorting of $d=5$, $d=10$ and $d=25$ spatial modes from the Fourier basis. Experimentally measured crosstalk matrices are shown in Figs.~\ref{Figure3}.a-b and return values of average sorting ability ranging from $93(3)\% $ ($d=5)$ to $25(10)\%$ ($d=25$). Moreover, it is essential to note that our approach can be used with any spatial mode basis. Figures~\ref{Figure3}.d-f show results of similar mode sorting experiments performed with modes randomly chosen within a set of LG modes of radial number $p=0$ and azimuthal number $\ell \in [\![-12;12]\!]$. Here we recalculated the matrix $P$ associated with the LG modes basis and the mode sorting phase masks using Eq.~\eqref{equ1} while using the same, unmodified TM. Figures~\ref{Figure3}.d-f show the measured crosstalk matrices and calculated phase masks (insets) used for sorting $d=5$, $d=10$ and $d=15$ modes, with sorting ability values between $82(3)\%$ to $15(6)\%$. 

Using the setup shown in Fig.~\ref{Figure1}.a, the experimental tests performed to characterise a given mode sorter are limited to sets of input modes that can be created by phase only modulation. To illustrate the versatility of our approach, we therefore simulated results of mode sorting using a random basis, in both amplitude and phase. A set of modes $\vec{R_i}$ with $i \in [\![1;25]\!]$ was selected from a numerically generated random complex hermitian unitary $1024\times1024$ matrix (see SI). This matrix was used as $P$ in Eq.~\ref{equ1} together with an experimentally measured TM $T$ to calculate the mode sorting phase masks. Propagation through the MMF was then numerically simulated by multiplying the phase shaped input fields by the experimentally measured TM. Results of crosstalk matrices and phase masks are shown in Figs.~\ref{Figure3}.g-h for $d=5$, $d=10$ and $d=25$ random modes, respectively, with average sorting ability values ranging from $97(2) \%$ to $45(17) \%$. These results confirm that our approach can be used to sort spatial modes from any arbitrary basis, independently of their complexity. 

\begin{figure}
\includegraphics[width=1 \columnwidth]{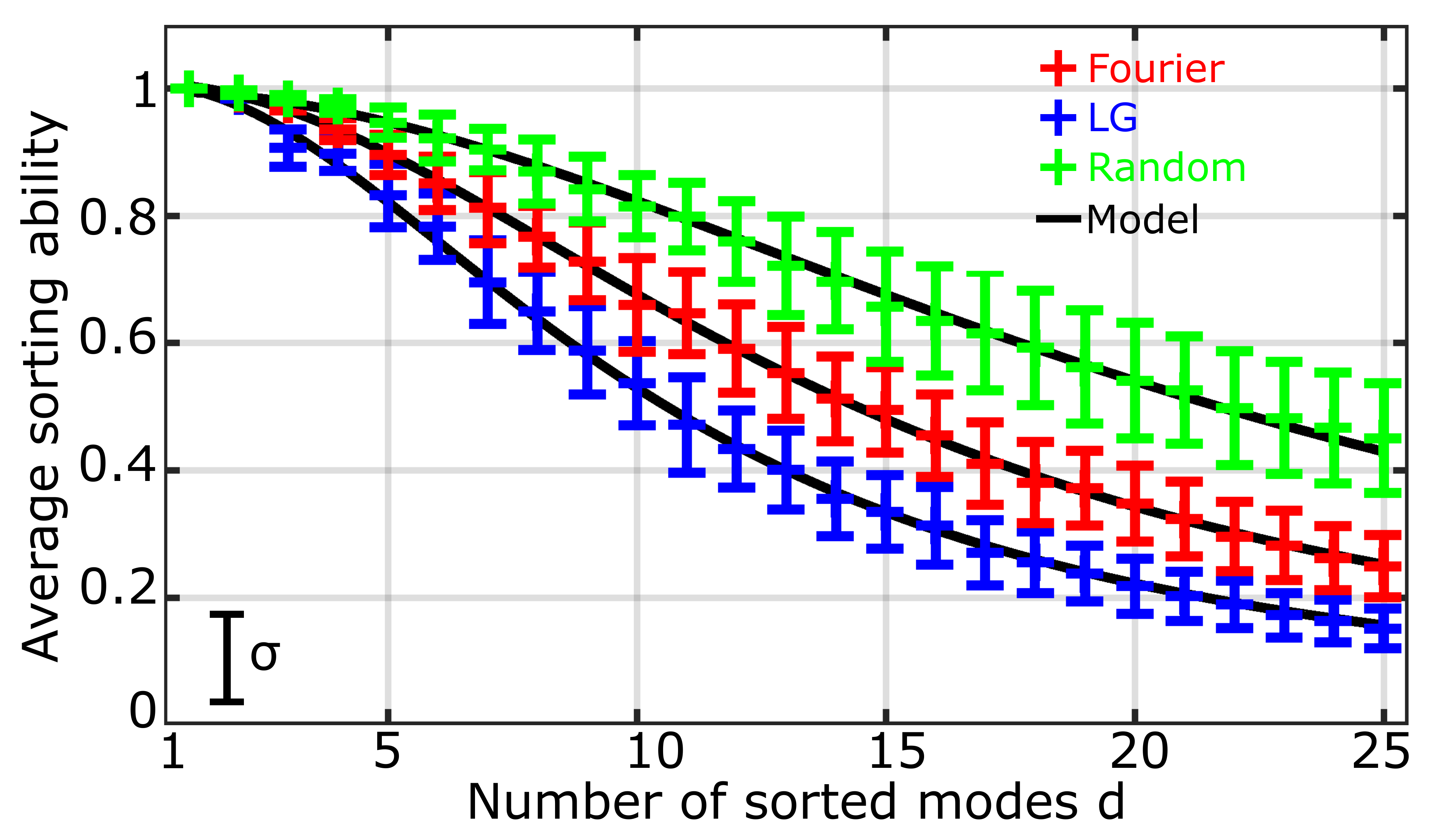} % this command will be ignored
\caption{\label{Figure4} Average sorting ability values for the Fourier (red), LG (blue) and random basis (green) in function of the number of modes $d$. Each value is measured by averaging over $10$ experiments performed with randomly chosen input modes and output positions. Results in the case of the random basis are obtained by simulating light propagation using an experimentally measured TM. The error bars represent the variations of the sorting ability values within the corresponding crosstalk matrices. Fitting models of the form $1/(1+Ad^2)+B$ are represented as solid black lines. Fitting processes return the parameters $A=0.005$ and $B=0.008$ (Fourier), $A=0.009$ and $B=0.008$ (LG) and $A=0.002$ and $B=-0.004$ (random), with coefficients of determination $r^2>0.997$. }
\end{figure}

Finally, a quantitative analysis of mode sorting performance is provided in Fig.~\ref{Figure4}. Values of average sorting ability are represented in function of the numbers of sorted modes $d$ for the Fourier (red), LG (blue) and random (green) basis. Experimental and simulated values are in very good agreement with a model of the form $1/(1+Ad^2)+B$, where $A$ and $B$ are two fitting parameters (see SI). These results show that the average sorting ability decreases with the increase of the number of sorted modes. We also observe that the variations of sorting abilities (error bars in Fig.~\ref{Figure4}) become larger for high $d$ values. However, it is important to note that the average focusing enhancement in our experiment is only 89, a value that can be improved by using a MMF supporting more modes (i.e. with a larger core diameter) and controlling more SLM macropixels. Improving the ability to focus light will decrease the value of the slope parameter $A$ and would enable to sort a larger number of modes with better sorting ability (see SI). 

In conclusion, we implemented an arbitrary spatial mode sorter in a multi-mode fibre using a TM-based wavefront shaping technique. Once the TM of the MMF has been experimentally measured, it is used it for sorting up to $25$ modes from a Fourier, an LG or a random basis. The sorting ability scales as $1/d^2$ with the number of sorted modes $d$. While an arbitrary mode sorting system would require $6d+1$ programmable phase screens to sort $d$ modes~\cite{lopez-pastor_arbitrary_2019}, our approach bypasses this constraint by harnessing the complex mixing process of a MMF using wavefront shaping, at the cost of a loss in overall efficiency. This loss of efficiency results from the compromise made to be able to sort spatial modes from arbitrary basis. Of course, if a mode sorting device only aims to operate on one specific type of mode (e.g. LG modes), it is more efficient to use the other implementations already reported in the literature~\cite{morizur_programmable_2010,lavery_refractive_2012,berkhout_efficient_2010,ruffato_compact_2018,mirhosseini_efficient_2013,fontaine_laguerre-gaussian_2019,fickler_full-field_2020}. In essence, we show that random mixing of light, usually considered as a drawback for imaging and communication, can be turned into an advantage for spatial mode sorting applications. Beyond mode sorting, our approach is also promising in communication schemes in which it is required to not only transport spatially multiplexed information but also to sort the information at the output, as for example in high-dimensional quantum communication schemes~\cite{valencia_unscrambling_2019,cao_distribution_2020,liu_multidimensional_2020}. \\
\\
\noindent \textbf{Aknowledgements.} D.F. acknowledges financial support from the UK Engineering and Physical Sciences Research Council (grants EP/T00097X/1 and EP/R030081/1) and from the European Union's Horizon 2020 research and innovation programme under grant agreement No 801060. H.D. acknowledges financial support from the EU Marie Skłodowska-Curie Actions (project 840958).\\

\bibliographystyle{apsrev4-1}
\bibliography{Biblio}

\newpage

\clearpage
\section*{Supplementary information} 

\subsection*{Details of the TM measurements and focusing process.}
\paragraph{TM measurement} The TM of the MMF is measured using the technique detailed in Ref.~\cite{popoff_measuring_2010}. At the input, the SLM is divided into $32 \times 32 = 1024$ macropixels composed of $16$ pixels each of size $8 \mu$m. At the output, optical field values are measured on $80 \times 40 = 3200$ camera pixels by phase-stepping holography using a non-modulated speckle as a reference. The SLM is a Holoeye Pluto NIR-II and the MMF is a $5$cm long $50 \mu$m core diameter graded index MMF from Thorlabs. 
\paragraph{Enhancement ratio} The enhancement ratio is defined as the ratio between the intensity at a target position on which light is focused and the average intensity before focusing~\cite{vellekoop_focusing_2007}. This ratio characterises the ability of our system to focus light through the MMF using the TM. In our experiment, we measures an average focusing enhancement value of $89$ (average value taken over $25$ focusing targets).
\paragraph{Approximation in equation~\eqref{equ2}} We consider $T$ as an $M \times N$ matrix composed of randomly distributed complex independent coefficients, with variance $\sigma_T^2$. The elements of $T T^\dagger$ can be written as:
\begin{itemize}
\item  Any off-diagonal element $[T T^\dagger]_{kl}$ ($k \neq l$) results from the complex sum of $N$ random phasors:
\begin{equation}
[T T^\dagger]_{kl} = \sum_{n=1}^N t_{ln} t_{kn}*
\end{equation} 
Therefore, $[T T^\dagger]_{kl}$ is also a random phasor with amplitude $|[T T^\dagger]_{kl}| = \sqrt{N} \sigma_T^2$ (i.e. random walk in the complex plane).
\item  Any diagonal element $[T T^\dagger]_{kk}$ results from the complex sum of $N$ real elements:
\begin{equation}
[T T^\dagger]_{kk} = \sum_{n=1}^N |t_{kn}|^2
\end{equation} 
Therefore, $[T T^\dagger]_{kk} = N \sigma_T^2$.
\end{itemize}
In conclusion, $T T^\dagger$ can be written as:
\begin{equation}
 T T^\dagger = \sigma_T^2 \left[ 1\!\!1 + \frac{H}{\sqrt{N}} \right]
\end{equation}
 where $H$ is a random matrix of complex coefficients with unity variance. Equation~\ref{equ2} is then only valid for $N \gg 1$, which is the case in our experiment ($N=1024$). Note that we can write $\sigma_T^2 = 1$ by normalising $T$ accordingly.

\subsection*{Details of the spatial input modes basis sets}
\begin{figure}
\includegraphics[width=1 \columnwidth]{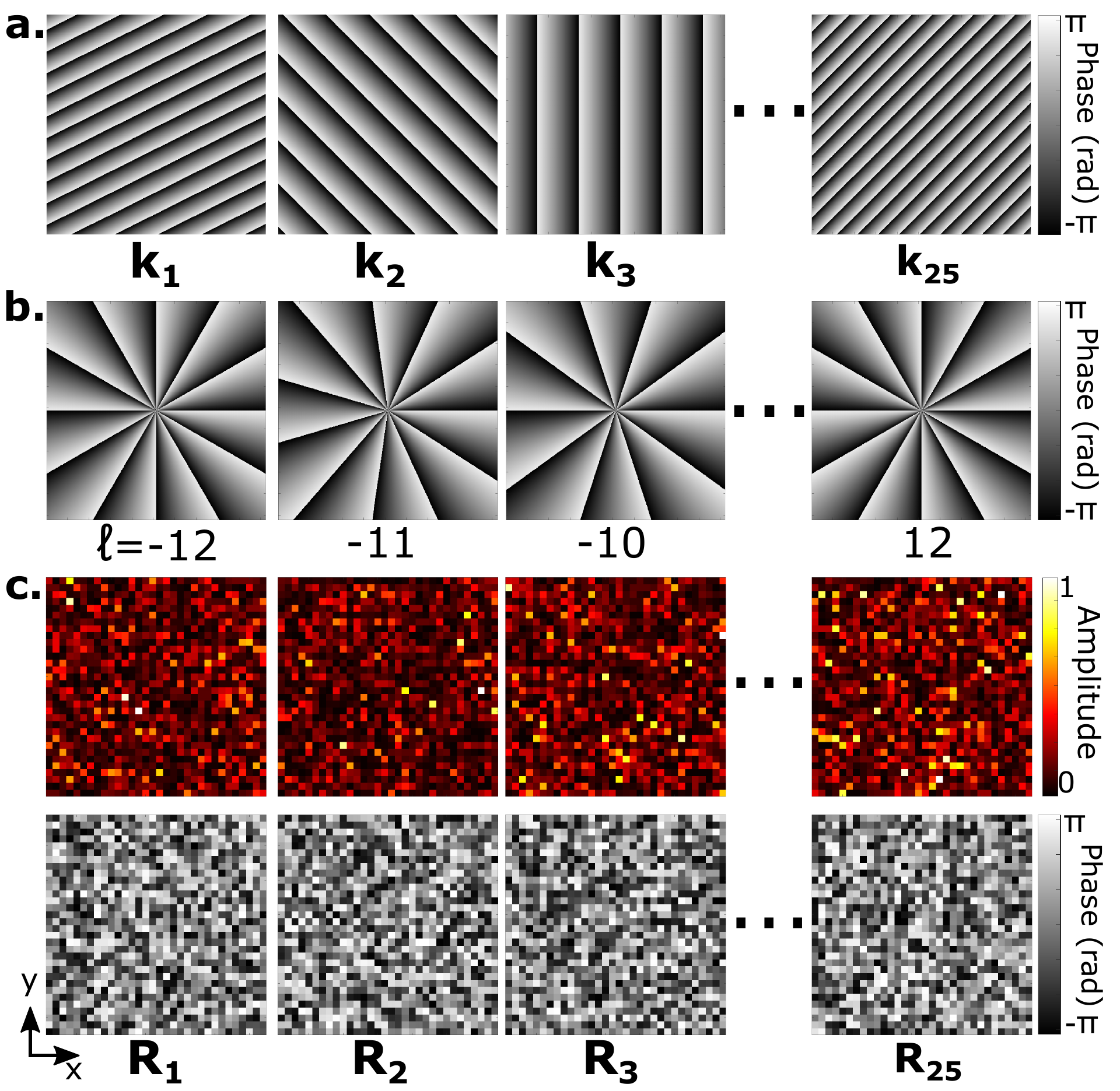} % this command will be ignored
\caption{\label{FigureSM} \textbf{a} Spatial phase component of $25$ input modes selected $\vec{k_i} $within the Fourier basis. \textbf{b} Spatial phase component of $25$ input modes with azimutal number $\ell_j \in [\![-12;12]\!]$ selected in the LG basis. \textbf{c}, Spatial amplitude and phase components of $25$ input modes $\vec{R_i}$ selected in the random basis. }
\end{figure}

\paragraph{Fourier basis.} After spatial discretization, an element $P_{ij}$ of the change of basis matrix $P$ associated to the Fourier basis is written:
 \begin{equation}
P_{ij} = e^{i \vec{k_j} \vec{r_i}}
\end{equation}
where $\vec{r_i}$ is the position of the $i^{th}$ macropixel of the SLM and $\vec{k_j}$ is the wave-vector associated to the $j^{th}$ input mode. In our experiment, we selected $25$ input modes with discrete wave-vectors $\vec{k_j} = (kx_j,ky_j)$ and with values $kx_j \in \{-2.9;-1.44;0;1.44;2.9\}.10^4$ $rad.m^{-1}$ and $ky_j \in \{-2.9;-1.44:0;1.44;2.9.\}.10^4$. The matrix $P$ is used in Eq.~\eqref{equ1} to calculate the mode sorting SLM phase mask. When performing the mode sorting experiments shown in Figs.~\ref{Figure2} and \ref{Figure3}.a-c, the phase patterns associated to the corresponding input mode are superimposed on top of the mode sorting phase mask on the SLM. These phase masks are shown in Fig.~\ref{FigureSM}.a. 
\paragraph{Laguerre-Gaussian basis.} After spatial discretization, an element $P_{ij}$ of the change-of-basis matrix $P$ associated with LG basis is written as:
 \begin{equation}
P_{ij} = e^{- \frac{|\vec{r_i}|.^2}{\omega}} e^{- i \ell_j \phi_i}
\end{equation}
where $|\vec{r_i}|$ and $\phi_i$ are the cylindrical coordinates of the $i^{th}$ macropixel of the SLM, $\omega \approx 1.7$mm is the waist of the collimated Gaussian beam illuminating the SLM and $\ell_j$ is the azimuthal number associated to the $j^{th}$ input mode. In our experiment, we selected $25$ inputs modes with $\ell_j \in [\![-12;12]\!]$. The matrix $P$ is used in Eq~\eqref{equ1} to calculate the mode sorting SLM phase mask. When performing the mode sorting experiments shown in Figs.~\ref{Figure3}.d-f, the phase patterns associated to the corresponding input modes are superimposed on top of the mode sorting phase mask on the SLM. These phase masks are represented in Fig.~\ref{FigureSM}.b.
\paragraph{Random basis.} The change-of-basis matrix $P$ associated with the random basis is a random complex unitary hermitian matrix of size $1024\times1024$ numerically generated by a computer. We selected $25$ inputs modes of this basis denoted $\vec{R_i}$ with $i \in [\![1;25]\!]$. The corresponding sub-matrix is used in Eq.~\eqref{equ1} to calculate the mode sorting SLM phase mask. Amplitudes and phases of the selected modes are shown in Fig.~\ref{FigureSM}.c. As explained in the main text, the results of shown in Figs.~\ref{Figure3}.g-i and \ref{Figure4} were obtained by numerically simulating the propagation of the random input modes using an experimentally measured TM.

\subsection*{Fitting model of sorting ability.}

\paragraph{Definition of the sorting ability.} The sorting ability $p_n$ associated with the $n^{th}$ input mode is calculated from the coefficient of the corresponding cross-talk matrix using the formula~\cite{fickler_custom-tailored_2017,fickler_full-field_2020}
\begin{equation}
p_n = \frac{I_{nn}}{\sum_{k=1}^d I_{kn}}
\end{equation}
where $I_{kn}$ is the coefficient of the crosstalk matrix linking input mode $n$ to the output position $k$. The average sorting ability $\bar{p}$ is therefore calculated by averaging over all the input modes sorting ability values:
\begin{equation}
\bar{p} = \sum_{n=1}^d p_n
\end{equation}
The associated standard deviation $\sigma_p$ is calculated using the formula:
\begin{equation}
\sigma_p = \sqrt{\frac{\sum_{n=1}^d (p_n-\bar{p})^2}{d}}
\end{equation}

\paragraph{Fitting model.} We build the fitting model of $\bar{p}$ (Fig.~\ref{Figure4}) on theoretical results already reported in~\cite{vellekoop_focusing_2007,popoff_measuring_2010}. We analyse separately the diagonal and off-diagonal coefficients of the crosstalk matrix:
\begin{itemize}
\item The SLM phase mask calculated using the TM and programmed onto the SLM to implement a $d$-dimensional mode sorting operation results from the superposition of $d$ phase masks, each mask being the phase pattern used for focusing light at a given position on the camera when the SLM is illuminated with a given input mode. Such a phase mask superposition is analogous to the phase mask superposition process used for focusing light through a complex system at multiple positions with an SLM illuminated by a single constant mode. In this latter case, it is demonstrated that the focusing intensity decreases as $1/d$, with $d$ being the number of target positions~\cite{vellekoop_focusing_2007,popoff_measuring_2010}. By analogy, we conclude that the diagonal coefficients of the cross-talk matrix scale as $I_{nn} \sim 1/d$.
\item Average intensity of the crosstalk matrix off-diagonal coefficients equals that of the grains in the output speckle pattern~\cite{vellekoop_focusing_2007}. This average intensity value is constant. Therefore, their sum scales as $d$.
\end{itemize}
We therefore conclude that:
\begin{equation}
\bar{p} = \frac{1}{d} \sum_{n=1}^d \left[ \frac{I_{nn}}{I_{nn} + \sum_{k\neq n} I_{kn}} \right] \sim \frac{1}{1+A d^2}
\end{equation}
where $A$ is a coefficient that depends on the focusing efficiency of our system, which includes the number of active macropixels of the SLM. The black curves shown in Fig.~\ref{Figure4} are obtained by fitting the experimental data with a model of the form $ \frac{1}{1+A d^2} + B$, where $\{A,B\}$ are the fitting parameters. 

\begin{figure}
\includegraphics[width=1 \columnwidth]{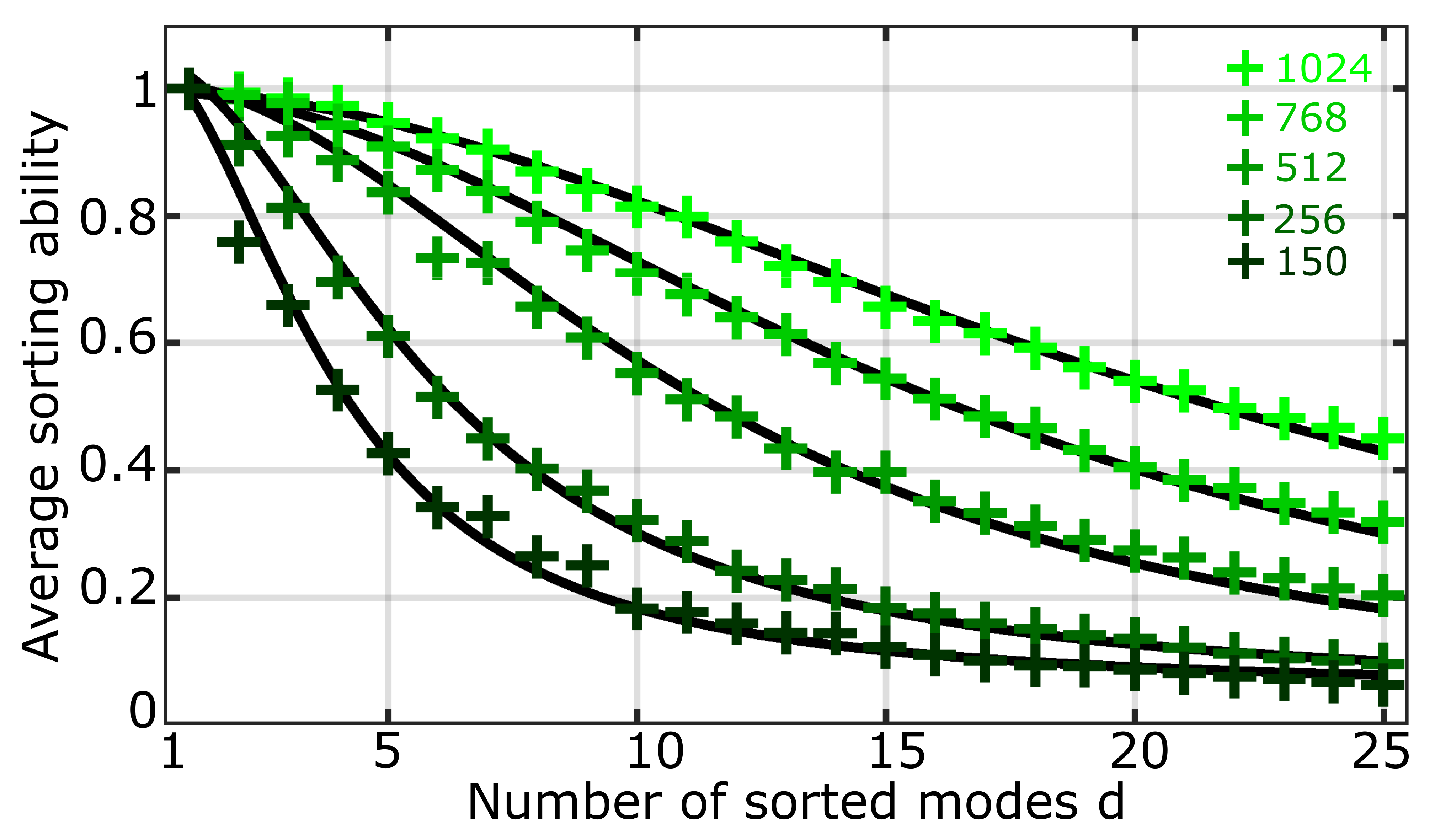} % this command will be ignored
\caption{\label{FigureSM1} The simulated average sorting ability values for random basis mode sorting are shown as a  function of the number of modes $d$ using a different number of active macropixels on the SLM. From top to bottom: the curve associated with $N=1024$ returns a coefficient of $A=0.002$, $N=768$ returns a coefficient of $A=0.004$, $N=512$ returns a coefficient of $A=0.008$, $N=256$ returns a coefficient of $A=0.03$ and $N=150$ returns a coefficient of $A=0.06$.   }
\end{figure}

\paragraph{Dependance of $A$ with the number of active macropixels on the SLM.}
The dependance of the parameter $A$ with the number of active macropixels on the SLM is investigated in Fig.~\eqref{FigureSM1}. This figure shows the results of five simulations of mode sorting in a random basis performed using different numbers of active macropixels on the SLM, ranging from $1024$ to $150$. We observe that the value of the coefficient $A$, and therefore the slope of the corresponding curve, increases when the number of active macropixels decreases. This confirms that the parameter $A$ directly depends on our ability to refocus light through the MMF.

\end{document}